\documentclass{elsart}

\usepackage{amssymb}
\usepackage{graphicx}
\usepackage{amsfonts}

\begin{document}

\begin{frontmatter}



\title{A Control Variate Approach for Improving Efficiency of Ensemble Monte Carlo\thanksref{ACK1}}
\thanks[ACK1]{Portions of this work, LA-UR-08-05399, were carried out at Los Alamos
National Laboratory under the auspices of the US National Nuclear Security Administration of the US Department of Energy. Tarik
Borogovac and Pirooz Vakili were supported in part by the National Science Foundation grants CMMI-0620965 and DGE-0221680.}

\author[TBad1,TBad2]{Tarik Borogovac\corauthref{cor1}}
\ead{tarikb@bu.edu} \corauth[cor1]{Corresponding author}
\author[TBad2]{Francis J. Alexander}
\author[PVad1]{Pirooz Vakili}

\address[TBad1]{Electrical and Computer Engineering Department, Boston University, 8 Saint Mary's St. Boston, MA 02215, USA}
\address[TBad2]{CCS-3, Los Alamos National Laboratory, MS-B256, Los Alamos, NM 87545, USA}
\address[PVad1]{Division of Systems Engineering \& Mechanical Engineering Department, Boston University, 15 Saint Mary's
St. Brookline, MA 02446, USA}

\begin{abstract}
In this paper we present a new approach to control variates for improving computational efficiency of Ensemble Monte Carlo. We
present the approach using simulation of paths of a time-dependent nonlinear stochastic equation. The core idea is to extract
information at one or more nominal model parameters and use this information to gain estimation efficiency at neighboring
parameters. This idea is the basis of a general strategy, called DataBase Monte Carlo (DBMC), for improving efficiency of Monte
Carlo. In this paper we describe how this strategy can be implemented using the variance reduction technique of Control Variates
(CV). We show that, once an initial setup cost for extracting information is incurred, this approach can lead to significant
gains in computational efficiency. The initial setup cost is justified in projects that require a large number of estimations or
in those that are to be performed under real-time constraints.
\end{abstract}

\end{frontmatter}

\section{Introduction}

The purpose of this paper is to present a novel approach for efficient estimation via the Monte Carlo (MC) method. The approach
is very broadly applicable but here, to present the main ideas, we narrow the focus to Ensemble Monte Carlo where estimation is
based on stochastically independent trajectories of a system. To illustrate, we use simulation of time-dependent nonlinear
processes for which Monte Carlo is a particularly general and powerful numerical method compared to available alternatives.
Time-dependent nonlinear processes are very general models used, among others, in statistical mechanics \cite{Bin1}, data
assimilation in climate, weather and ocean modeling \cite{Eve2}, financial modeling \cite{Glass1}, and quantitative biology
\cite{Wil1}. Hence developing efficient MC methods may significantly impact a wide range of applications.

A known weakness of MC is its slow rate of convergence. Assume $Y$ is a random quantity defined on paths of a process and let
$\sigma_Y$ denote its standard deviation. The convergence rate of MC for estimating the expected value of $Y$ is $\approx
\sigma_Y /\sqrt{n}$ where $n$ is the number of independent paths of the process. In general the canonical $n^{-1/2}$ rate of
convergence cannot be improved upon, hence, since the inception of the MC method, a number of variance reduction (VR) techniques
have been devised to reduce $\sigma_Y$ (see, \cite{HH64} for an early account and \cite{Glass1} and \cite{AsmGlynn} for more
recent discussions).

Most VR techniques lead to estimators
of the form
\[w_1Y_1+\cdots+w_nY_n,\]
i.e., a weighted average of the samples. These techniques prescribe (i) a recipe for selecting samples $Y_1, \cdots, Y_n$ and
(ii) a set of weights $w_1, \cdots,w_n$. To arrive at these prescriptions, one must rely on the existence of specific problem
features and the ability of the user of the method to discover and effectively exploit such features. This lack of generality has
significantly limited the applicability of VR techniques.

The point of departure of a new strategy, called DataBase Monte Carlo (DBMC), is to address this shortcoming and to devise
generic VR techniques that can be generically applied \cite{vak1}. All VR techniques bring additional information to bear on the
estimation problem, however, as mentioned above, this information is problem specific and relies on exploiting special features
of the problem at hand. By contrast, as will be clarified in this paper, DBMC adds a generic computational exploration phase to
the estimation problem that relies on gathering information at one (or more) nominal model parameter(s) to achieve estimation
efficiency at neighboring parameters. The advantage of this approach is its generality and wide applicability: it is quite easy
to implement and it can wrap existing ensemble MC codes. On the other hand, the computational exploration phase of the DBMC
approach may require extensive simulations and can be computationally costly. Therefore, the initial setup cost needs
justification. The setup cost may be justified in projects that involve estimations at many model parameters and/or in projects
where there is a real-time computational constraint. In the first type of project, the setup cost may lead to efficiency gain for
each subsequent estimation, and for a large enough number of subsequent estimations it can be easily justified. In projects with
a real-time constraint the setup cost is an off-line ``passive'' cost that can lead to estimates of significantly higher quality
(lower statistical error); the higher quality in many such projects more than justifies the setup cost.

In this paper we limit ourselves to presenting the implementation of the VR technique of Control Variates (CV) in the DBMC
setting (see \cite{vak1} for discussion of other VR techniques). The CV technique, which compared to the VR technique of
Importance Sampling is less utilized in computational physics, requires identifying a number of random variables called {\em
control variates}, say $X_1, \cdots, X_k$, that are correlated with $Y$ and have \emph{known} means. The correlation with $Y$
implies that $X_i$'s carry information about $Y$. The CV technique is a way of utilizing the information included in the controls
(their known means) to help with the estimation of the mean of variable $Y$. In the DBMC setting we assume that $Y=Y(\theta)$
depends on a model parameter $\theta$ and use $X_i=Y(\theta_i)$ where $\theta_i$'s are in a neighborhood of $\theta$
($i=1,\cdots,k$). In a departure from the classical CV technique, we use ``high quality'' estimates of $E[X_i]$ rather than
precise values of $E[X_i]$ to arrive at the controlled estimator of $E[Y]$. As we argue in this paper (and elsewhere
\cite{Boro1}) this departure allows for substantially broader choices of control variates and makes the CV technique
significantly more flexible and effective.

The DBMC method shares a similar intent as the well-known histogram reweighing method \cite{Fer1} from the Markov chain Monte
Carlo literature (e.g. \cite{Bin1}), but with a very different setting and implementation, and with broader applicability. For
example, it does not rely on having a Boltzmann distribution or $exp(-H/kT)$ structure. Given its generality, it has potential
applications, among others, in ensemble weather prediction, hydrological source location, climate and ocean, optimal control, and
stochastic simulations of biological systems.

The remainder of the paper is organized as follows. In section \ref{Prel} we discuss preliminaries, including the details of the
example numerical study -- the time-dependent Ginzburg-Landau (TDGL) equation -- as well as the method of control variates.
Estimation of mean outcomes of the TDGL equation over a range of temperatures is of interest, especially considering the large
difference in behavior below and above the coexistence curve. In section \ref{Methods} we describe the DBMC methodology and
motivation in a general context. Section \ref{Numerical} discusses the implementation and results of DBMC as applied to
estimation of quantities generated by the TDGL equation, and the results of that numerical study. We conclude in section
\ref{Conclusions}.

\section{Preliminaries}\label{Prel}
We present aspects of our approach and numerical results in the context of the time-dependent Ginzburg Landau (TDGL) model. It is
worth noting that this model is chosen for illustrative purposes only and we do not make use of any of its specific features.

\subsection{Time-Dependent Ginzburg Landau}

We use a canonical equation of phase-ordering kinetics \cite{Gul1,Ferreira1} the stochastic TDGL equation in two spatial
dimensions. This is written as
\begin{eqnarray}
\frac{\partial \phi({\bf x},t)}{\partial t} = D \Delta \phi({\bf x},t) - V'(\phi({\bf x},t))+ \eta({\bf x},t) \label{eq:TDGL}
\end{eqnarray}
where $\phi({\bf x},t)$ represents a local order parameter, e.g. a magnetization at point ${\bf x}=(x_1,x_2)^\top$ and time $t$
($^\top$ denotes transpose). The noise has mean zero and covariance $\langle \eta({\bf x},t) \eta({\bf x}',t')\rangle
=2\delta({\bf x}-{\bf x}') \delta(t-t')$. We choose a double-well potential $V(\phi) = - \frac{\theta}{2}\phi^2 +
\frac{\chi}{4}\phi^4$. As in \cite{Ferreira1} $\chi$ is a constant, and $\theta$ is a function of temperature such that a high
$\theta$ corresponds to a low temperature.

We use a discrete form of (\ref{eq:TDGL}) using a forward Euler-Maruyama stochastic integrator and a 5-point stencil for the
Laplacian (denoted $\Delta_L$) for simulation:
\begin{eqnarray}
\label{TDGLDIS}\nonumber \phi({\bf x}, t+\delta t)= \phi({\bf x}, t) &+& D \delta t\Delta_L\phi({\bf x},t)
 -\delta t[-\theta\phi({\bf x},t) + \chi\phi^3({\bf x},t)]\\
 \nonumber &+&
\sqrt{2 (\delta t/\delta x) } N({\bf x},t)
\end{eqnarray}
with time step $\delta t$ and lattice spacing $\delta x$, and where $N({\bf x},t)$ are independent and identically distributed
standard normal random variables for each space-time point $({\bf x},t)$. What follows applies to other discretization schemes as
well.

\subsection{Estimation problem}

To cover a broad range of estimation problems, we consider the estimation of quantities related to a specific space-time point,
quantities that are global (entire lattice at a particular time) and quantities that depend on the entire time evolution of the
system. Specifically, we consider the following representative quantities:

\begin{enumerate}
\item[(P1)] \emph{Point magnetization}: $\phi({\bf x}, t)$,
\item[(P2)] \emph{Total magnetization at a specific time $t$}: $\sum_{\bf x} \phi({\bf x},t)$,
 and
\item[(P3)] \emph{Total space-time magnetization}: $\sum_t\sum_{\bf x}\phi({\bf x},t)$.
 \end{enumerate}

The problem of estimating the expected value of any one of the above quantities can be represented by:
\[J(\theta)=E[Y(\omega; \theta))]\] where
$\omega$ is a vector of random numbers representing all the noise/uncertainty in a single complete path $\phi$ of the dynamics;
$\theta$ is the temperature related parameter; $Y(\omega;\theta)$ is the random sample of a quantity of interest (e.g., the magnetization
from a single sample path), and $E[\cdot]$ denotes expectation. Note that knowing the noise $\omega$ and parameter $\theta$
completely determines the path $\phi$ and the sample quantity of interest $Y(\omega;\theta)$.

\subsection{ The control variate technique}

Here we give a brief review of the classical control variate (CV) technique for variance reduction (see \cite{Glass1}
\cite{Rob1}).

Let $Y=Y(\omega;\theta)$, $J= E[Y]$. Assume $X_1, \cdots, X_k$ are random variables (called control variates) that are correlated
with $Y$ and assume their means $E[X_i]$ are known. Let ${\bf X}=(X_1, \cdots, X_k)^\top$, $E[{\bf X}]=(E[X_1], \cdots,
E[X_k])^\top$, and $\beta=(\beta_1, \cdots, \beta_k)^\top$. Then $Z$, defined below, is a controlled estimator of $E[Y]$
\[Z = Y + \sum_{i=1}^k\beta_i(X_i-E[X_i])= Y+
\beta^\top({\bf X}-E[{\bf X}])\] The estimator $Z$ uses information included in samples of the controls (the degree of their
deviation from their known means) to ``correct/adjust'' the estimator $Y$ and bring it closer to its unknown mean. This is the
key idea of CV. (Alternatively, $Z$ can be viewed as the fitted value of $Y$ when $Y$ is linearly regressed on variables $X_1,
\cdots, X_k$. In other words, $Z$ includes the part of the variation in $Y$ that cannot be ``explained'' by $X_i$'s.)

$Z$ is an unbiased estimator of $E[Y]$ for all vectors $\beta$. The coefficient vector that minimizes the
variance of $Z$ is:
\[\beta^o =\Sigma_{\bf X}^{-1}\Sigma_{{\bf X}Y}\]
where $\Sigma_{\bf X}$ is the $k\times k$ covariance matrix of ${\bf X}$ and $\Sigma_{{\bf X}Y}$ is the $k \times 1$ vector of
covariances of $Y$ and $X_i$'s. When $\beta^o$ is used, the variance of $Z$ is given by $(1-R^2)\sigma_Y^2$ where $R^2 =
\Sigma_{{\bf X}Y}^\top\Sigma_{\bf X}^{-1}\Sigma_{{\bf X}Y}/\sigma_Y^2.$ Therefore,
\begin{equation}
\frac{Var(Y)}{Var(Z)}=(1-R^2)^{-1}\label{eq: VRR_CV}
\end{equation} and hence $(1-R^2)^{-1}$ is precisely the theoretical degree of variance
reduction if the controlled estimator $Z= Y+ {\beta^{o\top}}({\bf X}-E[{\bf X}])$ is used to estimate $J$ as opposed to the crude
MC estimator $Y$, and it is called the Variance Reduction Ratio (VRR) statistic for control variates. Note that there is no upper
limit to the degree of achievable variance reduction since $R^2$ can potentially be very close to $1$ when the controls are
highly correlated with the estimation variable $Y$. In other words, the CV technique can potentially be very effective leading to
orders of magnitude of variance reduction.

In practice and in general, $\Sigma_{\bf X}$ and $\Sigma_{{\bf X}Y}$ (i.e. $\beta^o$) are not known exactly and need to be
estimated from samples of $X_i$'s and $Y$. Typically, $\beta^o$ is estimated from the same samples used to construct the
controlled estimator $Z$. While this practice adds some bias for small sample sizes, and thus makes the effective decrease in
estimator mean squared error not precisely equal to the variance reduction ratio $(1-R^2)^{-1}$, this bias converges to zero
faster than the standard error of $Z$. Thus, expending computational resources into generating separate pilot samples for
estimating $\beta^o$ is not considered to be justifiable. For an insightful and detailed discussion of the CV technique, see
\cite{Glass1}.

\subsection{Challenges in using the CV technique}
The critical task for using the CV technique is in finding effective controls. Once the controls are selected, the rest of the
procedure is fairly routine. An effective control, say $X$, needs to satisfy two requirements (to simplify the discussion we
consider a scalar control):
\begin{enumerate}
\item[(R1)] $X$ needs to be correlated with $Y$, and
\item[(R2)] $E[X]$ needs to be available to the user, i.e., known.
\end{enumerate}
The main barrier to finding effective controls is the second requirement, namely the requirement of a known mean $E[X]$. A
modification of the CV technique called Biased Control Variate (BCV) reduces the burden of requirement (R2) by allowing for a
good approximation of $E[X]$ when $E[X]$ cannot be evaluated analytically \cite{Schm1}. While BCV lowers the requirement barrier
and expands the range of available choices for effective controls, it nonetheless limits its potential scope by implicitly
assuming an analytic path to arriving at the approximate value. As we describe in the next section, in the DBMC approach we turn
the second requirement into a computational task; in other words, we use statistical estimation to obtain a good estimate of
$E[X]$. Therefore, barrier (R2) is completely removed and the range of choices of controls is dramatically expanded. The relevant
question now becomes whether the computational investment in estimating $E[X]$ pays enough dividends to make the investment
worthwhile.

\section{DBMC \& Control Variate} \label{Methods}

The starting point of the DBMC approach is the observation that in many parametric estimation settings, including in the example
considered in this paper, quantities $Y(\theta)$ and $Y(\theta')$ are highly correlated when the same random input $\omega$ is
used to generate them and when $\theta$ and $\theta'$ are close\footnote{A similar observation is the basis for the histogram
re-weighting methods: ``from a simulation at a single state point (characterized in an Ising model by choice of temperature T and
magnetic field H) one does not gain information on properties at that point only, but also in the neighboring region,''
(\cite{Bin1}, page 116)}. This suggests using control variates $X_i = Y(\theta_i)$, $i=1,\cdots, k$, when estimating $Y(\theta)$
where $\theta_i$'s are ``close'' to $\theta$.

While we have identified potentially effective controls, we do not have sufficient information about them, i.e.,
$J(\theta_i)=E[X_i]$ is not known and needs to be evaluated. This brings us to the second feature of the DBMC method that
corresponds to its initial computational information gathering/setup stage. This stage corresponds to statistical estimation of
$J(\theta_i)$. Details are given below.

\subsection{DBMC + CV algorithm}

The DBMC approach consists of a setup stage and an estimation stage.

\subsubsection{Setup stage}

The DBMC setup phase involves generating a ``large'' number of input random vectors $\omega$ and obtaining ``high quality''
estimates of $J(\theta_i)$. Let $DB= \{\omega_1, \omega_2, \cdots, \omega_N\}$ ($N$ ``large'') denote a large set of random
inputs. This set represents the {\em database}. Given the database, the averages of the controls are precisely calculated. A
schematic of this stage is given in Figure \ref{fig: DBMC setup}.

\begin{figure}[h]
\rule{\columnwidth}{1pt}
\begin{enumerate}
\item For $j=1, \cdots, N$
\begin{enumerate}
\item Generate $\omega_j$ according to the distribution of the inputs;
\item For $i=1,\cdots,k$
\begin{enumerate}
\item Simulate the path $\phi(\omega_j; \theta_i)$
\item Evaluate the value of the control $X_i(\omega_j)= Y(\omega_j; \theta_i)$\\
\end{enumerate}
\end{enumerate}
\item For $i=1,\cdots,k$
\begin{enumerate}
\item Find $J_{DB}(\theta_i)$, the average of the $i$th control on the datebase, as \[J_{DB}(\theta_i)= \frac{1}{N}\sum_{j=1}^N Y(\omega_j; \theta_i)\]
\end{enumerate}
\end{enumerate}
\rule{\columnwidth}{1pt} \caption{DBMC setup stage}\label{fig: DBMC setup}
\end{figure}

\subsubsection{Estimation stage}

To estimate $J(\theta)$, at a $\theta$ close to $\theta_i$'s ($i=1,\cdots, k$) select a ``small'' sample (say of size $n\ll N$)
uniformly from the database. For each sample $\omega_j$ re-simulate the equation using $\omega_j$ and $\theta$ to obtain
$Y(\omega_j; \theta)$. For these samples the values of the controls $X_i(\omega_j)$ are available in the database. Using these
evaluate a controlled estimate of $J(\theta)$. A schematic version of these steps is given in Figure \ref{fig: DBMC estimation}.

\begin{figure}[h]
\rule{\columnwidth}{1pt}
\begin{enumerate}
\item For $j=1, \cdots, n$
\begin{enumerate}
\item Select $\omega_j$ uniformly from the database;
\item Simulate the path $\phi(\omega_j; \theta)$;
\item Evaluate the estimation variable $Y(\omega_j; \theta)$.\\
\end{enumerate}
\item Find the controlled estimator of $J(\theta)$:
\begin{equation}\label{DBCV}
\widehat{J}_{cv}(\theta)=\frac{1}{n}\sum_{j=1}^n[Y(\omega_j; \theta)+\sum_{i=1}^k \beta_i^o(X_i(\omega_j)-J_{DB}(\theta_i))]
\end{equation}
\end{enumerate}
\rule{\columnwidth}{1pt} \caption{DBMC estimation stage}\label{fig: DBMC estimation}
\end{figure}

\subsection{Implementation choices}

There are two general schemes for implementation of our CV approach: (I1) corresponding to what is described above, requires
storing simulation inputs $\{\omega_j\}$ and outputs $\{X_i(\omega_j)\}$ in a database for later resampling; (I2) does not
utilize resampling, so there is no storage of data beyond recording the calculated control means. Both implementations are
feasible, the first is preferable in most cases; the second may be preferred in some cases. We elaborate below.

Implementation (I1).
\begin{itemize}
\item The database of random inputs, i.e., $\omega_j$'s, are either directly stored or enough information about them (e.g.
input seeds of a pseudo-random number generator) is stored to be able to regenerate $\omega_j$'s precisely.
\item The $k$ paths corresponding to $\theta_i$, $i=1, \cdots, k$, i.e., $\phi(\omega_j;\theta_i)$ are generally simulated ``in
parallel'' as elements of a random vector, $\omega_i$, are progressively generated.
\item For each random input, say $\omega_j$, the value of the controls, $X_i(\omega_j)$, $j=1, \dots, N$, $i=1,\dots,k$ are
stored.
\end{itemize}
Implementation (I2).
\begin{itemize}
\item Once the setup stage is completed, the only values stored are the ``high quality'' estimates of the means of $J(\theta_i)$'s,
i.e., the $k$ values $J_{DB}(\theta_i)$, $i=1,\cdots,k$.
\item At the estimation phase, $n$ random input vectors $\omega_j$,
$j=1, \dots,n$, are generated {\em anew}; paths at $\theta$ and $\theta_i$, $i=1,\cdots, k$, are simulated using new random
inputs and for each path $X_i(\omega_j)$ and $Y(\omega_j;\theta)$ are calculated; finally, using these values, the controlled
estimator is evaluated.
\end{itemize}

\subsection{Statistical properties \& computational efficiency}

The promise of the approach is the following: by anchoring estimation via CV at a few high quality estimates (at
$\theta_1,\cdots, \theta_k$), it is possible to obtain high quality estimates at other locations in the parameter space (at other
$\theta$) with far fewer samples. The actual statistical properties of the resulting estimators, and the computational efficiency
of generating them, reflect choices made in implementing each given problem. For example, how much computation should be
``invested'' in the exploration phase, and which points $\theta_i$ in the parameter space should be explored are two important
questions that need further investigation. Such choices generally involve problem dependent tradeoffs, and we leave them to
future studies.

Instead, the analysis that follows is meant to provide a general and qualitative understanding of the statistical properties,
computational efficiency and the tradeoffs involved. The discussion is as general as possible, but consistent with the numerical
study described in section \ref{Numerical}, where such implementation choices were made utilizing only a basic familiarity with
the problem. For further discussion, see \cite{Boro1}.

\subsubsection{Statistical properties}

We give the analysis for implementation (I1). In other words, assume we are re-sampling from the database. Analysis of
implementation (I2) shows similar estimator statistical properties.

To simplify the discussion consider a single control, say $X_1$. Let $J(\theta)=E[Y]$, $J(\theta_1)=E[X_1]$, $\sigma_Y^2={\mbox
Var}(Y)$, $\sigma_{X_1}^2={\mbox Var}(X_1)$. Assume a database of input variables are generated and let $Y^*$ and $X^*_1$ be
random variables corresponding to $Y$ and $X_1$ that are generated by re-sampling (uniformly, with replacement) from the
database. Let $J^*(\theta)$, $\sigma_{Y^*}^2$, $J^*(\theta_1)$, $\sigma_{X^*_1}^2$ denote the means and variances of the
re-sampled variables $Y^*$ and $X^*$.

Conditioned on the database, the controlled estimator is exactly the classical CV estimator and all results from classical CV
apply. For example, for any scalar $\beta$, $Z^*=Y^*+ \beta(X^*-J^*(\theta_1))$ is an unbiased estimator of $J^*(\theta)$,
$J^*(\theta_1)$ is known, and the optimal $\beta^o$ is what is prescribed by classical CV if we take all random variables as
those defined on the database. A measure of variance reduction due to using a controlled estimator is
\begin{equation}
VRR=\frac{\sigma^2_{Y^*}}{\sigma^2_{Z^*}}\label{eq: VRR_CV*}
\end{equation}

We use the controlled estimator $Z^*$ as an estimator for $J(\theta)$. Assume optimal $\beta^o$ is used to define $Z^*$ and
assume $E[Z^*]=J^*(\theta)$ \footnote{i.e. we ignore the low order bias that results from the typical CV procedure of estimating
the optimal $\beta^o$ e.g. \cite{Glass1}, not to be confused with the resampling bias discussed in this section}. In general
$J^*(\theta) \neq J(\theta)$. Therefore, $Z^*$ is a biased estimator of $J(\theta)$ where the bias is introduced by sampling from
the database, i.e., from $Y^*$, as opposed to from $Y$. We have some probabilistic assessment of this bias and we can reduce it
by increasing the size of the database. Specifically, for this bias we can obtain an approximate $1-\alpha$ probability
confidence interval:
\[P(|J^*(\theta)- J(\theta)| \leq \frac{z_{\alpha/2}\sigma_Y}{\sqrt{N}})\approx 1-\alpha\]
where $z_{\alpha/2}$ is the $1-\alpha/2$ quantile from the standard normal distribution. In other words, with high probability
the bias is of the order of $O(N^{-1/2})$. We assume that for large $N$ the bias is sufficiently small to be disregarded and that
we can focus on $VRR$ in (\ref{eq: VRR_CV*}) as the key measure of computational gain in using the controlled estimator to
estimate $J(\theta)$.

\subsubsection{Computational efficiency}
Generating the above large database, as we pointed out earlier, corresponds to an initial ``setup'' cost. Let $C$ be the
computational cost of generating a sample of $Y(\theta)$. This cost involves generating an $\omega$, simulating the path, and
evaluating $Y(\omega, \theta)$. A reasonable assumption for many problems is that this cost is about the same for all $\theta$
and $\omega$. Then the set-up cost of generating the database and obtaining averages of the controls is approximately $N\times k
\times C$. Let $VRR(\theta)$ denote the variance reduction ratio at $\theta$, i.e., the ratio of the variance of an uncontrolled
sample and that of a controlled sample at $\theta$. Then, the statistical error of a controlled estimator based on $n$ samples is
approximately the same as that of $n \times VRR(\theta)$ samples of an uncontrolled estimator. Thus, the ratios of the
computational costs of the two estimators (to arrive at the same statistical accuracy) is $(n \times VRR(\theta) \times C)/(n
\times C) = VRR(\theta)$. Therefore, $VRR(\cdot)$ can serve as a measure of benefit of the DBMC approach.

The setup cost of the DBMC approach can be justified in two types of applications. The first type are those applications that
require solving many instances of the estimation problem, at many $\theta$'s. If the total number of instances is sufficiently
large, and some variance reduction is achieved on the average on those instances, then the large fixed set-up cost can be dwarfed
by the total computational savings from the many estimations. The second type are real-time applications where the setup cost can
be viewed as an off-line cost enabling significant efficiency gains in the critical task of real-time estimation. Typically, the
``cost'' of delay in such real-time estimation is higher and not merely computational, justifying even a much larger
computational effort off-line.

\section{Numerical results} \label{Numerical}

The numerical results in this section are intended to give a qualitative illustration of the efficiency gains that can be
achieved using the DBMC approach. Specifically, we estimate the variance reduction that can be achieved over regular (crude)
sampling, when estimating the three quantities of interest (a point magnetization, total magnetization at a specific time $t$ and
the total time-space magnetization) at a range of the parameter $\theta$. Our choices of the size of the database, number of
samples used for estimation, range of parameter values, and the controls are simply for illustration purposes. However, we expect
that the numerical results are, qualitatively, quite representative.

We simulate the TDGL dynamics on a $40\times40$ lattice (lattice spacing $\delta x_k=1$, $k=1,2$) with fixed $\chi=1$. On each
path, we evolve the system for a total of $5000$ time steps ($\delta t=0.01$) which is sufficient for the system to exhibit
behavior that is specific to its temperature region. The critical point for this system is $\theta_c=1.265$ \cite{Ferreira1}, and
our parameter range of interest ($1.0$ to $1.5$) extends to both sides of that critical point.

To build a database, we simulate $N=2^{14}=16384$ paths and evaluate point magnetization, total magnetization at a specific time,
and total space-time magnetization at two nominal values of $\theta$, $1.2$ and $1.35$.

For each quantity of interest, we consider three control variate estimators. The first two estimators, CV1.2 and CV1.35, use
single controls corresponding to $\theta=1.2$ and $\theta=1.35$, respectively. We chose to anchor our estimators at those two
nominal values for $\theta$ because they are located on opposite sides of the phase transition line $\theta_c$. The third
estimator, CV2C, uses both controls simultaneously.

\begin{figure}[hbt]
\includegraphics[scale=0.8]{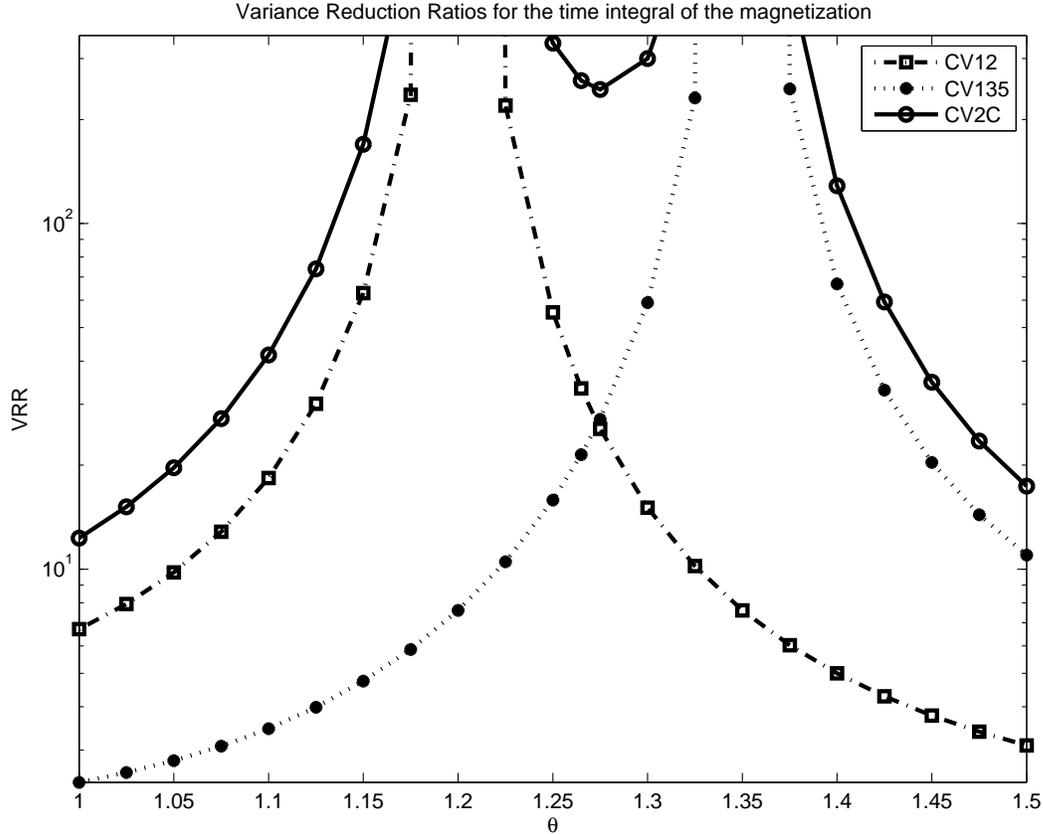}
\caption{Variance reduction ratios of the estimators of the space-time integral of the magnetization, $\sum_{\bf x}\sum_t
\phi({\bf x},t)$, over a range of values of $\theta$ (log scale). \label{fig: VRR_tim_log}}
\end{figure}

\begin{figure}[hbt]
\includegraphics[scale=0.8]{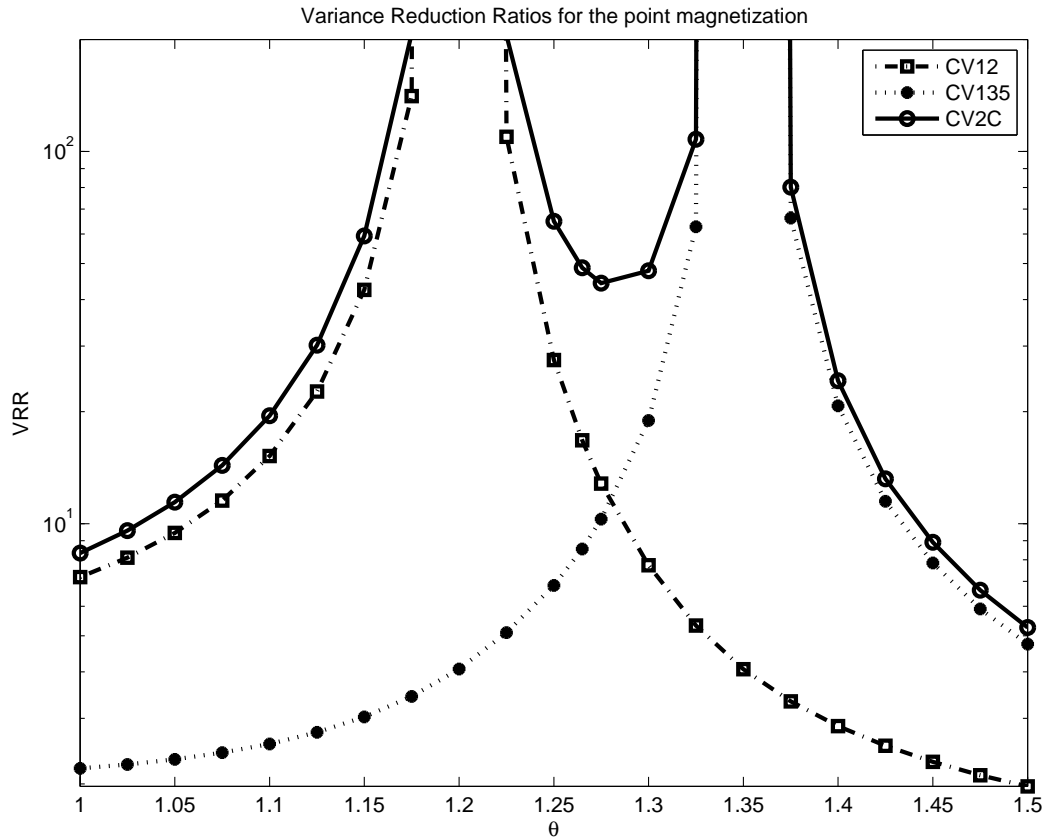}
\caption{Variance reduction ratios of the estimators of the point magnetization over a range of values of $\theta$, in log scale.
\label{fig: VRR_point_log}}
\end{figure}

We use $n=2^8=256$ samples for crude and CV estimators. To estimate the variance of these estimators, following the micro-macro
simulation approach (see, e.g., \cite{Schm3}), we use $40$ independent macro simulations consisting of $256$ independent micro
simulations. We obtain variance estimates from each macro simulation and average the resulting $40$ values to obtain an overall
variance estimate. We report the ratios of the variance estimates (crude/controlled, as in (\ref{eq: VRR_CV*})) as $VRR$. A
sampling of VRR results for the total space-time magnetization (problem P3) is given in Table \ref{tab:table1} and the
corresponding graph is given in Figure \ref{fig: VRR_tim_log}. The graph for point magnetization (problem P1) are given in Fig.
\ref{fig: VRR_point_log}, and the results for the total magnetization at a time $t$ (problem P2) are quite similar and are
excluded.

\begin{table}[hbt]
\caption{\label{tab:table1}Variance reduction ratios of the estimators applied to space-time integral of the magnetization,
$\sum_{\bf x}\sum_t \phi({\bf x},t)$, at several values of $\theta$.}
\begin{tabular}{lccccccccc}
 Estimator$\backslash\theta$ &1.150 &1.175 &1.225 &1.250 &1.265 &1.300 &1.325 &1.375 & 1.400\\
\hline CV1.2 &63 &236 &219 &55 &33 &15 &10 &6 &5  \\
CV1.35 &5 &6 &11 &16 &21 &59 &231 &245 &67  \\
CV2C &170 &709 &947 &332 & 259 &300 &761 &513 &129  \\
\end{tabular}
\end{table}

Based on these results, we draw the following conclusions:
\begin{itemize}
\item Controlled estimators produce dramatic variance reduction for
parameter values very close to the nominal parameters and substantial variance reduction at values moderately close to the
nominal.
\item For all the estimation problems, adding the second control consistently improves performance, in some cases
leading to substantial reduction in variance (compared to single controls). Of course, by incorporating information from points
on both sides of the critical temperature, CV2C is expected to give better coverage than either of the single control estimators.
However, CV2C does better than either of the single control estimators even in their own regions, which suggests that each
control provides relevant information to the estimation problem in the opposite region.
\item VRR values for the total space-time
magnetization are somewhat larger than those for the point and total magnetization at a specific time $t$ -- we expect this to be
true more generally for path integrals when compared with values at specific time instances.
\end{itemize}

\section{Conclusions}\label{Conclusions}

In this paper we described a new strategy, DataBase Monte Carlo (DBMC), for improving computational efficiency of Ensemble Monte
Carlo. For a specific time-dependent nonlinear dynamics we showed that the approach can lead to significant efficiency gains for
a range of estimation problems. Our selection of the controls has been ad-hoc and for illustration purposes. Further work is
required to better understand the options available and the computational tradeoffs involved. To this end, our current research
is focused on (i) derivation of more specific guidelines for the selection of effective control variates, (ii) implementation of
the DBMC strategy in conjunction with other variance reduction techniques, for example, stratification and importance sampling,
and (iii) application of the method in some specific domains, for example, estimation problems in geophysical fluids and
biochemical systems.

\bibliographystyle{elsart-num}

\end{document}